\documentclass[nofootinbib, aps, prd, a4paper, 10pt, superscriptaddress, eqsecnum, showkeys]{revtex4-2}
\usepackage[a4paper, top=2cm, bottom=2cm, left=2.5cm, right=2.5cm]{geometry}

\usepackage{amsmath}
\usepackage{amsfonts}
\usepackage{amsthm}
\usepackage{bm}
\usepackage{mathrsfs}
\usepackage{braket}

\usepackage{graphicx}
\usepackage{booktabs}

\usepackage{xcolor}
\usepackage[pdfusetitle]{hyperref}
\hypersetup{colorlinks = true, allcolors = blue}

\usepackage{orcidlink}

\begin{document}


\title{Scale-dependent Dark Matter in a Multi-component Cosmological Framework: Distinct Components Governing Galactic and Cosmological Dynamics}
\author{V.K. Oikonomou\orcidlink{0000-0003-0125-4160}}
\email{voikonomou@gapps.auth.gr} \affiliation{Department of
Physics, Aristotle University of Thessaloniki, Thessaloniki 54124,
Greece} \affiliation{Center for Theoretical Physics, Khazar
University, 41 Mehseti Str., Baku, AZ-1096, Azerbaijan}

\begin{abstract}
In this work we generalize the concept of scale-dependent equation
of state dark matter from galactic to cosmological scales. At
galactic scales, the scale-dependent equation of state which
successfully describes dark matter dominated galaxies has a nearly
isothermal equation of state of the form $P(r)=K(r)\rho(r)$, where
the entropy function $K(r)$ is radius dependent and slowly varying
with the galactic radius. Using appropriate theoretical arguments,
the successful galactic scale dark matter model is generalized to
have a universal equation of state of the form $\langle
P(\chi)\rangle =\langle \rho(\chi)\rangle \langle
\sigma_{DM}^2\rangle$, where $\chi$ is the scale of the system
under study and it can describe cosmological or galactic scales
and $\sigma_{DM}$ is the characteristic velocity dispersion of
dark matter at the relevant scales. This results to the entropy
function behaving universally as $K(\chi)=\langle
\sigma_{DM}^2\rangle$. We investigate the cosmological
implications of scale-dependent equation of state dark matter, and
we conclude that this scale-dependent equation of state dark
matter cannot be the sole component of dark matter in the
Universe. Also the second aim of this work is to provide a
sufficient explanation for the question why $K_0$ changes for each
galaxy, where $K_0$ is the value of $K(r)$ at intermediate and
large radii from the core of the galaxy. Furthermore, we introduce
the concept of having distinct dark matter components which
cluster differently, and thus controlling differently the dynamics
of the Universe at various scales. Scale-dependent equation of
state dark matter can control the galactic scales, while another
component can dominate at cosmological scales and thus control the
Universe at cosmological scales.
\end{abstract}

\maketitle

\section{Introduction}

To date, the existence of dark matter (DM) is not confirmed
experimentally, but is rather confirmed observationally by its
cosmological scale and galactic scale effects. The situation is
somewhat similar to the situation of black holes in 1960 when
these Einstein field equations solutions were nothing but
mathematical predictions of general relativity. Through the years
though, more and more dynamical effects in astrophysical systems
pointed to the existence of black holes, with the most important
the trajectories of stars in the center of our galaxy. Thus we
knew that black holes are real, and we knew that they caused
strong gravitational effects in their surroundings but we did not
have direct proof of a black hole until 2019. Now, the same
applies with DM, we know its out there but no direct experimental
verification of DM matter has been done, to date though. But DM
controls galactic dynamics and the very own large scale structure
is due to DM creating the first strong gravitational potentials in
the places of the primordial perturbations, and in these
potentials would ordinary matter collapse and virialized to create
the galactic structures we see in the Universe. The absence of DM
would make the cosmic microwave background (CMB) impossible to
consistently explain it, gravitational lensing would also have no
explanation, and the rotation curves of galaxies would be hard
also to explain without modifying gravity. DM was firstly observed
by Zwicky in 1933, who used the mass-to-light ratio from the Coma
cluster and had it compared to the mass-to-light ratio of spiral
galaxies. His conclusion was that there was at least 400 times
more unseen mass in the Coma cluster.

The benchmark model for cosmology and astrophysics to date is the
$\Lambda$-Cold-Dark-Matter ($\Lambda$CDM) model, and one of its
basic ingredients is cold DM. Although overall its quite
successful for nearly 3 decades, it is challenged with tensions to
present date, both at cosmological
\cite{DiValentino:2020naf,Dai:2020rfo,He:2020zns,Nakai:2020oit,Agrawal:2019dlm,Yang:2018euj,
Ye:2020btb,Vagnozzi:2021tjv,Desmond:2019ygn,Hogas:2023pjz,OColgain:2018czj,Vagnozzi:2019ezj,
Krishnan:2020obg,Colgain:2019joh,Vagnozzi:2021gjh,Lee:2022cyh,Krishnan:2021dyb,Ye:2021iwa,Ye:2022afu,Verde:2019ivm,Menci:2024rbq,Adil:2023ara,Reeves:2022aoi,Ferlito:2022mok,
Vagnozzi:2021quy,DiValentino:2020evt,DiValentino:2019ffd,DiValentino:2025sru,Valletta:2025bgu,Montani:2025nmz,Fazzari:2025mww,Montani:2025jkk,Schiavone:2024heb,Montani:2024pou,Montani:2024ntj,Escamilla:2024xmz,Montani:2023ywn,Montani:2023xpd}
and galactic levels. At the galactic level, there are persistent
problems that need to be explained like the cusp-core problem in
dwarfs, low-luminosity and low surface brightness spirals, the
too-big-to-fail problem and the diversity problem
\cite{Tulin:2017ara}. So the galaxies that impose challenges on
the $\Lambda$CDM, and thus on the cold DM perspective, are the
DM-dominated galaxies, so irregular dwarfs, low-luminosity and low
surface brightness spirals. For these galaxies, it seems that
self-interacting DM can soften the problems, compared to cold DM.
In this spirit, in some recent works
\cite{Oikonomou:2026zhu,Oikonomou:2026seq,Oikonomou:2025bsi,Oikonomou:2026vkp}
we introduced and analyzed the concept of scale-dependent equation
of state (EoS) DM, in which DM obeys a generalized EoS of the form
$P(r)=K(r)\rho^{\gamma(r)}$, which is a sort of polytropic EoS.
Polytropic EoS DM is a sort of self-interacting DM, and both
polytropic and self-interacting DM are frequently used in the
literature
\cite{Slepian:2011ev,Saxton:2014swa,Alonso-Alvarez:2024gdz,Kaplinghat:2015aga,Ahn:2004xt,Saxton:2012ja,Saxton:2010jk,Saxton:2016ozz,Arbey:2003sj,
Heikinheimo:2015kra,Wandelt:2000ad,Spergel:1999mh,Loeb:2010gj,Ackerman:2008kmp,Goodman:2000tg,Arbey:2006it,Yang:2025ume,Abedin:2025dis},
see also an interesting work on non-linear regimes of DM
\cite{Nunes:2026wfg}.

In this work we aim to investigate how the galactic scale nearly
isothermal scale-dependent EoS DM behaves at cosmological scales.
Using theoretical arguments, we end up to the more universal
relation $\langle P(\chi)\rangle =\langle \rho(\chi)\rangle
\langle \sigma_{DM}^2\rangle$ which results to $K(\chi)=\langle
\sigma_{DM}^2\rangle$, where $\chi$ is the scale of the structure
under study and $\sigma_{DM}$ is the characteristic velocity
dispersion of DM at the scale of the structure under
consideration. We examine the cosmological implications of a
scale-dependent EoS and we show that the concept of
scale-dependent EoS DM has a consistent cosmological description
too. However, this scale-dependent EoS DM cannot be the only DM
component in the Universe, since it can lead to observational
inconsistencies, so we also introduce the concept of having two or
more components of DM, which cluster in different ways. Thus some
of the DM components can contribute only at cosmological scales,
while scale-dependent DM can contribute and affect significantly
only the galactic scales.

\section{Motivation for a Scale-dependent EoS for DM at Galactic Scales}

In this section we shall provide a concrete description of the
scale-dependent EoS DM perspective and discuss its main
achievements at galactic scales. For details see Refs.
\cite{Oikonomou:2026zhu,Oikonomou:2026seq,Oikonomou:2025bsi}. The
scale-dependent self-interacting DM perspective for DM is based on
the following generalized EoS,
\begin{equation}\label{eq:eos_variable}
P(r)=K(r)\left(\frac{\rho}{\rho_{\star}}\right)^{\gamma(r)}\, ,
\end{equation}
which effectively describes self-interacting DM due to the
presence of a non-trivial pressure. The new ingredient in this
sort of EoS is the term $K(r)$, which is effectively an entropy
related term, which captures the non-trivial dynamical properties
of DM at galactic scales. This entropy term can in principle model
in a feasible way the effective interactions and virilization of
DM at galactic scales as was shown in Refs.
\cite{Oikonomou:2026zhu,Oikonomou:2025bsi}. For simplicity in this
work we shall assume that the scale-dependent EoS we shall
consider is the following,
\begin{equation}\label{eosmain}
P(r)=K(r)\rho(r)^{\gamma(r)}\, ,
\end{equation}
so the entropy function $K(r)$ has units $(km/sec)^2$. Regarding
some successful profiles for scale-dependent DM, the following
profile,
\begin{align}\label{iniDMpres}
    \gamma(r) &= \gamma_0 - \delta_\gamma\,\tanh\!\left(\frac{r - r_\gamma}{2.0\mathrm{Kpc}}\right), \\
    K(r) &= K_0\,\left(1 + \frac{r}{r_c}\right)^{-p}\, ,
\end{align}
with
\begin{align}\label{modelparametervalues}
  & \gamma_0=1.0001\\ \notag &
  \delta_\gamma=1.2\times 10^{-9}\\ \notag &
  r_c=0.5\,\mathrm{Kpc}\\ \notag &
  p=0.01\\ \notag &
  r_\gamma=1.5 \,\mathrm{Kpc}
\end{align}
resulted to successfully reproducing the galactic rotation curves
of all irregular galaxies, intermediate low-surface brightness and
low-luminosity spirals from the SPARC data, while simultaneously
reproducing empirically the Baryonic and Canonical Tully-Fisher
relations for all the above types of galaxies. Thus a
scale-dependent EoS profile of the form (\ref{eosmain}) can
describe successfully galaxies and solve the cusp-core problems of
galactic dynamics. These galaxies are DM-dominated, thus
scale-dependent EoS can successfully describe these galaxies from
a dynamical point of view. There is a reason why DM profiles of
the sort (\ref{eosmain}) with $K(r)$ having the form
(\ref{iniDMpres}) and this was explained in Refs.
\cite{Oikonomou:2026zhu,Oikonomou:2026seq,Oikonomou:2025bsi}. In
this section we shall discuss the reasons in order to motivate the
use of scale-dependent EoS and to further discuss its
generalization at cosmological scales, but also to put its
specialization to each galaxy via the parameter $K_0$ or generally
via the function $K(r)$. The complete set of differential
equations that one has to solve for the scale-dependent DM and
which describe the hydrodynamic equilibrium, is the following,
\begin{align}\label{differentialequationsmainset}
\frac{dP}{dr} &= -\rho(r)\frac{G M(r)}{r^2}, \\
P(r) &= K(r) \rho(r)^{\gamma(r)}, \\
\frac{dM}{dr} &= 4\pi r^2 \rho(r)\, .
\end{align}
The circular velocity for stars in a galactic system is defined as
follows,
\begin{equation}
v_\phi^2(r) = \frac{G\,M(r)}{r}. \label{eq:v_circ}
\end{equation}
\begin{figure}[h!]
\centering
\includegraphics[width=20pc]{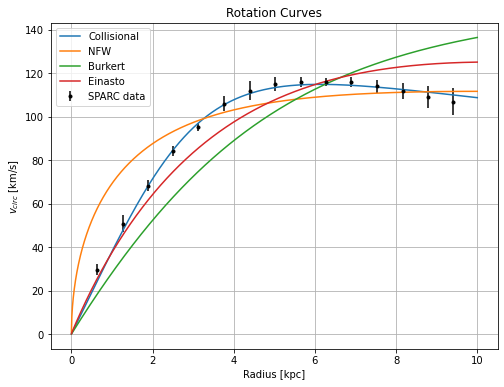}
\caption{The predicted rotation curves of the model
(\ref{iniDMpres}) for the galaxy UGC05986 versus the SPARC data.
We included the optimized curves for the NFW model, the Burkert
model and the Einasto model.}\label{UGC05986}
\end{figure}
For the scale-depended profile (\ref{iniDMpres}), from the set of
free parameters of the profile, the most important which play a
crucial role for the phenomenological viability of the model are
$K_0$ and $\rho_0$. The rest of the free parameters
(\ref{modelparametervalues}) are the same for all galaxies, but
only $K_0$ and $\rho_0$ change from galaxy to galaxy and were
fitted in Ref. \cite{Oikonomou:2025bsi} in order to comply with
the SPARC data. Now the important observation is the following,
from the parameter values (\ref{modelparametervalues}) it is
obvious that the EoS for the scale-dependent DM is nearly
isothermal at the core and the inner halo regions, because
essentially $\gamma \simeq 1$, thus,
\begin{equation}
P(r) \sim K(r)\,\rho(r)\, .
\end{equation}
For the profile $K(r)$ we used, namely,
\begin{equation}\label{krprofile}
K(r)= K_0\,\left(1 + \frac{r}{r_c}\right)^{-p}
\end{equation}
since $K(r)$ is very slowly-varying, for medium and large
distances $r$ from the core, we have approximately,
\begin{equation}\label{kapprox}
K(r)\sim K_0\, ,
\end{equation}
since $K(r)$ is slowly varying, and thus $K_0$ is rendered an
effective entropy constant. The phenomenological viability of the
model is due to this exact feature, that the entropy at large
radii is effectively constant, but in the inner parts of the
galaxy it contributes dynamically, changing its value as the
radius varies, thus fitting perfectly the rotation curves of
DM-dominated galaxies. Using
Eqs.~(\ref{differentialequationsmainset}) and (\ref{eosmain}) in
the large $r$ limit, with $K(r)$ approximated as in Eq.
(\ref{kapprox}), we get,
\begin{equation}
\frac{1}{\rho}\frac{d\rho}{dr} = -\frac{G\,M(r)}{K_0\,r^2}\, .
\end{equation}
\begin{figure}[h!]
\centering
\includegraphics[width=20pc]{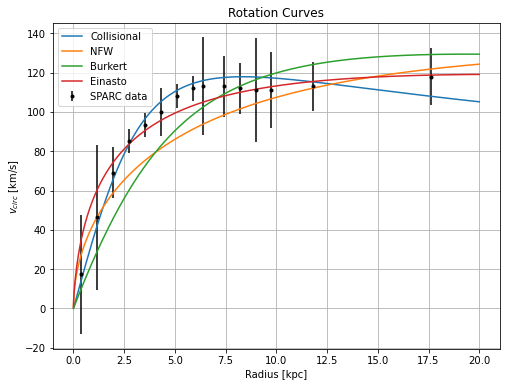}
\caption{The predicted rotation curves of the model
(\ref{iniDMpres}) for the galaxy F568-V1 versus the SPARC data. We
included the optimized curves for the NFW model, the Burkert model
and the Einasto model.}\label{F568-V1}
\end{figure}
If we assume a quasi-isothermal sphere, we have approximately,
\begin{equation}\label{rhoquasiisothermal}
\rho(r)\sim r^{-2}\, ,
\end{equation}
and thus,
\begin{equation}
M(r)\sim r, \quad \rightarrow \quad v_\phi^2(r)=\frac{GM(r)}{r} =
\text{constant}.
\end{equation}
This is an important behavior, since it shows that flat rotation
curves for DM dominated galaxies can be obtained naturally using a
scale-dependent EoS, by employing solely hydrodynamic equilibrium
and by using an entropy function $K(r)$ that effectively becomes a
constant $K(r)\sim K_0$. This constant value of the entropy
function at large radii affects medium and large radii, thus
yielding nearly flat rotation curves at these distances, but it
does not affect the core of the galaxy, thus providing an elegant
resolution to the cusp-core problem of DM-dominated galaxies. The
importance of a constant entropy function $K(r)$ at large radii is
essential in order to have a nearly isothermal DM density at
hydrostatic equilibrium and at large distances,
\begin{equation}
\rho(r)\sim r^{-2}\, ,
\end{equation}
and therefore one can obtain,
\begin{equation}
4\pi\int_0^r \rho(r')r'^2,dr' \propto r\, ,
\end{equation}
and from the combination of the above, we get,
\begin{equation}
\frac{GM(r)}{r} \simeq {\rm constant}\, .
\end{equation}
Thus the most successful scale-dependent DM profile is the one
which captures two features, it yields significantly fast
$K(r)\sim K_0$ and at the same time it yields an appropriate
behavior $K(r)$ near the core of the galaxy to solve the cusp-core
problem and also in order to fit optimally the galactic rotation
curves of DM-dominated galaxies.

As we already mentioned, the values of the entropy parameter $K_0$
and of the galactic density $\rho_0$ were fitted in Ref.
\cite{Oikonomou:2025bsi} in order to provide the rotation curves
of the SPARC galaxies. The values of $K_0$ and $\rho_0$ change
from galaxy to galaxy, and although it is natural to understand
why the galactic central density $\rho_0$ is distinct and unique
to each different galaxy, the reason why the value of $K_0$ that
changes for each galaxy was not discussed in Ref.
\cite{Oikonomou:2025bsi}. The aim of this work is to provide a
sufficient explanation for the question why $K_0$ changes for each
galaxy, and also to discuss how the scale-dependent EoS behaves at
cosmological and not galactic scales. This will also answer the
question regarding the behavior of $K(r)$ at galactic scales.

In order to show how the model (\ref{iniDMpres}) succeeds to fit
the SPARC data for DM dominated galaxies, we shall present some
galaxies and their rotation curves and the predictions of the
model (\ref{iniDMpres}) for these galaxies. We also include the
best fits for the NFW model, the Burkert model and the Einasto
model. We consider first the galaxy UGC05986 which is a
late-type/irregular spiral, so it is DM-dominated, of the sort
that the model (\ref{iniDMpres}) optimally succeeds to produce the
galactic rotation curves. For the galaxy UGC05986 we choose the
central density to be $\rho_0=8\times
10^7$$M_{\odot}/\mathrm{Kpc}^{3}$ and $K_0=1.5\times10^{3}$$
(km/sec)^2$ and the rest of the parameters chosen as in Eq.
(\ref{modelparametervalues}). In Fig. \ref{UGC05986} we plot the
predictions of the model (\ref{iniDMpres}) for the galaxy
UGC05986, versus the SPARC data for this galaxy.

\begin{figure}[h!]
\centering
\includegraphics[width=20pc]{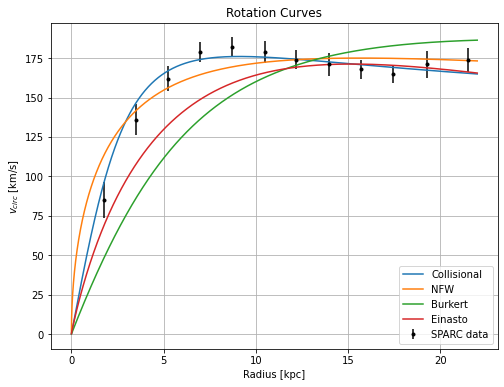}
\caption{The predicted rotation curves of the model
(\ref{iniDMpres}) for the galaxy NGC4088 versus the SPARC data. We
included the optimized curves for the NFW model, the Burkert model
and the Einasto model.}\label{NGC4088}
\end{figure}

Now let us consider the galaxy F568-V1 which is a faint late-type
low-surface-brightness disk galaxy, of the sort that the model
(\ref{iniDMpres}) better fits. For the galaxy F568-V1 we choose
the central density to be $\rho_0=7.8\times
10^7$$M_{\odot}/\mathrm{Kpc}^{3}$ and
$K_0=5.5\times10^{3}$$(km/sec)^2$ and the rest of the parameters
chosen as in Eq. (\ref{modelparametervalues}). In Fig.
\ref{F568-V1} we plot the predictions of the model
(\ref{iniDMpres}) for the galaxy UGC05986, versus the SPARC data
for this galaxy. Finally, let us present the galaxy NGC4088 which
is an intermediate spiral galaxy. For this galaxy we choose
$\rho_0=2.1\times 10^8$$M_{\odot}/\mathrm{Kpc}^{3}$ and
$K_0=1.25\times 10^{4}$$(km/sec)^2$ and the rest of the parameters
chosen as in Eq. (\ref{modelparametervalues}). In Fig.
\ref{NGC4088} we plot the predictions of the model
(\ref{iniDMpres}) for the galaxy UGC05986, versus the SPARC data
for this galaxy.
All the sample galaxies we chose to present are fitted perfectly
by the model (\ref{iniDMpres}) and all are DM-dominated, or simply
the baryons do not strongly affect the dynamics of the rotation
curves.

Apart from the model (\ref{iniDMpres}), we developed a full
analytic scale-dependent EoS DM model, studied in detail in Ref.
\cite{Oikonomou:2026zhu,Oikonomou:2026vkp}, which remarkably can
fit perfectly and reproduce the galactic rotation curves of
galaxies with significant cores, thus solving the cusp-core
problem for these galaxies. Specifically, the model of Ref.
\cite{Oikonomou:2026zhu,Oikonomou:2026vkp} can actually fit
perfectly the galactic rotation curves of the galaxy UGCA320,
which is also known as DDO161. This galaxy was studied recently in
Ref. \cite{refugca320} and it was shown that conventional DM
profiles, like the NFW and similar cannot reproduce the galactic
rotation curves. We showed in
\cite{Oikonomou:2026zhu,Oikonomou:2026vkp} and we also demonstrate
here that the scale-dependent EoS DM profile can reproduce the
galactic rotation curves of this galaxy, when gas, bulge and
baryonic contributions are also taken into account. A few words
for the model of Ref. \cite{Oikonomou:2026zhu,Oikonomou:2026vkp}
for completeness.

\begin{figure}[h!]
\centering
\includegraphics[width=20pc]{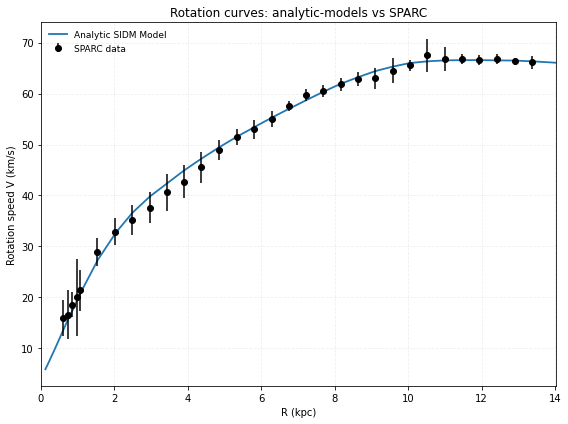}
\caption{The optimized rotation curves for the galaxy UGCA320
(\ref{ScaledependentEoSDM}), and the extended SPARC data for the
galaxy DDO161/UGCA320. We also included the rotation curves for
the other components of the galaxy, that is, the gas, the disk
velocities, the bulge (where present) and the scale-dependent EoS
DM model.} \label{extendedDDO161}
\end{figure}

For this model,the scale-dependent (radius dependent) EoS of DM is
of the form $P(r)=K(r)\left(\frac{\rho}{\rho_{\star}}\right)$,
with the specific energy density $\rho(r)$ and also the entropy
function $K(r)$ having analytic forms of the following form,
\begin{align}\label{ScaledependentEoSDM}
      \rho(r) &=\frac{\rho_0}{\left(
1+\frac{r^2}{\alpha^2}\right)^{\frac{5}{2}}}, \\
    K(r) &= \frac{K_0}{\left(
1+\frac{r^2}{\alpha^2}\right)^{\frac{1}{2}}}\, ,
\end{align}
with $\rho_\star=1$$M_\odot\,{\rm Kpc}^{-3}$ for simplicity. The
model is a two-parameter model with $K_0$ and $\rho_0$ being the
only free parameters, while the parameter $\alpha$ can be deduced
by the relation,
\begin{equation}\label{parameteralpha}
K_0= \frac{2 \pi  \alpha ^2 G \rho _0}{9 \rho _*}\, .
\end{equation}
The profile (\ref{ScaledependentEoSDM}) analytically satisfies the
following set of hydrodynamic equilibrium equations,
\begin{align}\label{differentialequationsmainset}
\frac{dP}{dr} &= -\rho(r)\frac{G M(r)}{r^2}, \\
P(r) &= K(r) \left(\frac{\rho(r)}{\rho_{\star}}\right)^{\gamma(r)}, \\
\frac{dM}{dr} &= 4\pi r^2 \rho(r)\, .
\end{align}
For the rotation curves, we shall also take into account the other
galactic contributions and specifically, the gas, the bulge and
disk components of the baryons. The total rotation curve velocity
is thus,
\[
V_{\text{total}}^2(r) = V_{\text{disk}}^2(r) +
V_{\text{bulge}}^2(r) + V_{\text{gas}}^2(r) + V_{\text{DM}}^2(r)\,
.
\]
With regard to the physical units, the pressure has units,
\begin{equation}
[P] = [\rho][v^2] = M_\odot\,{\rm Kpc}^{-3} \, ({\rm km/s})^2\, ,
\end{equation}
and also,
\[
P(r) = K(r)\frac{\rho(r)}{\rho_\star},
\]
hence,
\begin{equation}
[K(r)] = [P] = M_\odot\,{\rm Kpc}^{-3} \, ({\rm km/s})^2\, .
\end{equation}
Now let us consider the galaxy UGCA320 which was recently studied
in Ref. \cite{refugca320}. The galaxy UGCA320, also known as
DDO\,161, is a dwarf galaxy which is gas-rich and has
morphological type Sm, and it is usually classified as a dwarf
irregular. Its distance is $D =
6.03^{+0.29}_{-0.21}\,\mathrm{Mpc}$ and it forms an isolated pair
together with UGCA\,319. Active star formation and a significant
neutral hydrogen content characterize this galaxy. In Fig.
\ref{extendedDDO161} we showcase the combined rotation curves for
the galaxy UGCA320, where we included all the galactic components
of the galaxy, along with the scale-dependent EoS DM. As it can be
seen in Fig. \ref{extendedDDO161}, the galactic rotation curves
are generated perfectly.
In Table \ref{evaluationextendedDDO161} we describe which
optimized values for $K_0$ and $\rho_0$ can produce the galactic
rotation curves for the galaxy DDO161/UGCA320 and the reduced
$\chi^2_{red}$ value.
\begin{table}[h!]
\centering \caption{Parameter Values of the Scale-dependent DM
model analytic model of Eq. (\ref{ScaledependentEoSDM})  for the
Galaxy DDO161/UGCA320.}
\begin{tabular}{lc}
\hline
Parameter & Value  \\
\hline
$\chi^2_{red}$ & 0.202594 \\
$ml_{\text{disk}}$ & 0.8271 \\
$ml_{\text{bulge}}$ & 0.5529 \\
$K_0$ ($M_{\odot} \,
\mathrm{Kpc}^{-3} \, (\mathrm{km/s})^{2}$) & 1549.15   \\
$\rho_0 $ ($M_{\odot}/\mathrm{Kpc}^{3}$) & $4.03068 \times 10^6$   \\
\hline
\end{tabular}
\label{evaluationextendedDDO161}
\end{table}
Thus the scale-dependent EoS DM models can produce perfectly the
rotation curves of even observationally intriguing galaxies,
solving the cusp-core problem optimally.

\section{Scale-dependent DM EoS at Galactic and Cosmological Scales: A Unified Description}

In the spirit of the previous section, the behavior of the nearly
isothermal scale-dependent DM EoS at galactic scales will be,
\begin{equation}\label{eosmaincosmo}
P(a)=K(a)\rho(a)\, ,
\end{equation}
where the characteristic galactic local scale $r$ is replaced by
the scale factor $a$ of the Friedmann-Robertson-Walker metric,
\begin{equation}\label{FRW}
\mathrm{d}s^2=-\mathrm{d}t^2+a(t)^2\sum_i \mathrm{d}x_i^2\, .
\end{equation}
The crucial element for creating a link between the global
(\ref{eosmain}) and the local versions (\ref{eosmaincosmo}) of the
scale-dependent EoS is the characteristic velocity dispersion of
DM and of the nearly isothermal EoS it obeys. This is natural to
understand because nature cannot realize that the physical scale
actually changes from galactic to cosmological scales, but what
effectively changes is the characteristic DM velocity, and indeed
this changes even between galaxies. So in general, the temperature
depended EoS is,
\begin{equation}\label{temperaturedependendEoS}
P=\rho k_B T_{DM}/m_p\, ,
\end{equation}
where $m_p$ is the mass of the DM particle, $T_{DM}$ is the DM
characteristic temperature and $k_B$ is Boltzmann's constant. The
relation between the temperature and the characteristic velocity
distribution is,
\begin{equation}\label{temperatureandcharacteristicvelcoty}
T_{DM}=\frac{m_p\sigma_{DM}^2}{k_B}\, ,
\end{equation}
where $\sigma_{DM}$ is the characteristic DM velocity dispersion
which measures the random velocity of particles $u$ relative bulk
velocity $v$. This easily follows from the isotropic velocity
distribution tensor $\sigma_{ij}^2=\sigma_{DM}^2\delta_{ij}$ where
$\sigma_{DM}^2=\frac{1}{3}\langle |v-u|^2 \rangle$. Then from
kinetic theory, the kinetic energy with random particle motion is
$E_{kin}=\frac{1}{2}m_p \langle |v-u|^2 \rangle$, and for an
isotropic non-relativistic distribution, the kinetic theory
definition of the temperature is $E_{kin}=\frac{3}{2}k_B T$, thus
the relation (\ref{temperatureandcharacteristicvelcoty}).
Combining (\ref{temperaturedependendEoS}) and
(\ref{temperatureandcharacteristicvelcoty}) we get,
\begin{equation}\label{globalScaleDependentEoS}
P=\rho \sigma_{DM}^2\, .
\end{equation}
This is an important equation, since it is actually this relation
that changes from global scales to local scales. So the actual
scale-dependent EoS DM which describes both the global and local
scale-dependent profiles is,
\begin{equation}\label{globalScaleDependentEoSaverage}
\langle P(\chi)\rangle =\langle \rho(\chi)\rangle \langle
\sigma_{DM}^2\rangle \, .
\end{equation}
The averages run over the characteristic scale $\chi$ which may
describe the cosmological scales $\chi=a(t)$ or the local galactic
scales $r$. What actually determines the scale is the
characteristic velocity dispersion $\sigma_{DM}$. Thus, the
function $K(\chi)$ is related to the characteristic velocity
dispersion as,
\begin{equation}\label{characteristicvelocityandKr}
K(\chi)=\langle \sigma_{DM}^2\rangle\, ,
\end{equation}
and thus this is a rather universal relation.

Let us focus on galactic scales and we shall show why the actual
value of $K_0$ in Eq. (\ref{kapprox}) changes from galaxy to
galaxy. Recall that $K_0$ is the asymptotic value of $K(r)$ of
each galaxy at large distances from the center of each galaxy.
Consider that we have the galaxy ``i'' with radius $R_i$ and mass
$M_i$ which virializes at a redshift $z_i$ with virilization
velocity dispersion $\sigma_{DM_i}$. In the outer halo we have
$\sigma_{DM_i}$,
\begin{equation}\label{outveri}
\sigma_{DM_i}^2\sim \frac{G\,M_i}{R_i}\, .
\end{equation}
The mean density inside a virialized halo is,
\begin{equation}\label{meanvirdensity}
\bar{\rho}_i=\frac{3M_i}{4\pi
R_i^3}=\Delta_{vir}\rho^i_{crit}(z_i)\, ,
\end{equation}
where $\Delta_{vir}$ is the virial overdensity relative to the
critical density, and also $\rho^i_{crit}(z_i)$ is,
\begin{equation}\label{rcriti}
\rho^i_{crit}(z_i)=\frac{3H^2(z_i)}{8\pi G}\, ,
\end{equation}
so we have,
\begin{equation}\label{auxvir}
R_i^3=\frac{2GM_i}{\Delta_{vir}H^2(z_i)}\, .
\end{equation}
Thus combining the above we obtain,
\begin{equation}\label{rvirprefinal}
R_{vir}=\left(\frac{2G M}{\Delta_{vir}}
\right)^{1/3}H^{-2/3}(z_i)\, ,
\end{equation}
hence for an appropriately fixed virial overdensity we have.
\begin{equation}\label{rvirfinal}
R_{vir}\sim M^{1/3}H^{-2/3}(z_i)\, ,
\end{equation}
thus in view of Eq. (\ref{rvirfinal}), Eq. (\ref{outveri})
becomes,
\begin{equation}\label{finalvirializedrelation}
\sigma_{DM_i}^2\sim M_i^{2/3}H^{2/3}(z_i)\, ,
\end{equation}
hence each galaxy has a characteristic velocity dispersion
$\sigma_{DM_i}$ which depends both on the virial halo mass and
also on the Hubble rate $H(z_i)$ at the time that the halo
virialized. Therefore, each galaxy has a distinct characteristic
velocity dispersion and in view of Eq.
(\ref{characteristicvelocityandKr}), and Eq.
(\ref{characteristicvelocityandKr}), at galactic scales we have,
$K_{0_i}=\sigma_{DM_i}^2\sim M_i^{2/3}H^{2/3}(z_i)$, hence each
galaxy has a distinct value $K_{0_i}$ for the entropy parameter
$K_0$. This explains why the optimal $K_0$ which brings
phenomenological success in fitting the rotation curves of each
galaxy, varies from galaxy to galaxy.

Now the scale-dependent EoS of Eq.
(\ref{globalScaleDependentEoSaverage}) at cosmological scales has
distinct impact on the function $K(r)$ and the physical
consequences of scale-dependent EoS DM. Indeed, at cosmological
scales, the characteristic velocity dispersion of DM redshifts as,
\begin{equation}\label{scaledependedecosmo}
\sigma_{DM}\sim a^{-1}\, ,
\end{equation}
where $a(t)$ is the scale factor. Therefore, due to relations
(\ref{globalScaleDependentEoSaverage}) and
(\ref{characteristicvelocityandKr}), at galactic scales the
entropy function $K(\chi)$ will redshift as,
\begin{equation}\label{galacticentropyfunction}
K(a)=K_0\, a^{-2}\, .
\end{equation}
Also the coarse-grained DM EoS parameter
$w_{DM}=\frac{P_{DM}}{\rho_{DM}}$ will redshift as,
\begin{equation}\label{eoseparametercosmo}
w_{DM}=\frac{\sigma_{DM}^2}{c^2}=w_{DM_0}a^{-2}\, ,
\end{equation}
or written in terms of the redshift,
\begin{equation}\label{redshiftewedos}
w_{DM}=w_{DM_0}(1+z)^2\, ,
\end{equation}
where $w_{DM_0}$ is the value of the DM EoS parameter at present
day and also $c$ is the speed of light. In addition, the DM sound
speed for cosmological scales easily follows,
\begin{equation}\label{soundspeedcosmol1}
c_s^2=\frac{\partial P}{\partial \rho}=w_{DM}c^2\, .
\end{equation}
The continuity equation for this sort of self-interacting DM is,
\begin{equation}\label{contequation}
\dot{\rho}_{DM}+3 H\left(1+w_{DM}(a) \right)\rho_{DM}=0\, ,
\end{equation}
so by integrating we get,
\begin{equation}\label{integratedcont}
\rho_{DM}=\rho_{DM_0}e^{3\int_0^z\frac{1+w_{DM}(z')}{1+z'}\mathrm{d}z'}\,
,
\end{equation}
hence by using relation (\ref{redshiftewedos}) we finally obtain,
\begin{equation}\label{finalevoscaledependeneoscosmo}
\rho_{DM}(z)=\rho_{DM_0}\left(1+z\right)^3\,e^{3\frac{w_{DM_0}}{2}\left((1+z)^2-1
\right)}\, .
\end{equation}
When $w_{DM_0}=0$, the standard redshift for COLD DM is recovered.

Thus, in the context of the scale-dependent EoS DM, the
characteristic velocity dispersion is quantifying the entropy
parameter $K(\chi)$ which is present both at galactic and at
cosmological scales. This entropy parameter must be considered as
a coarse-grained variable, like the velocity dispersion of DM. At
cosmological scales it receives a universal value which is of the
order $K_0\sim \mathcal{O}(1)$$(km/sec)^2$ while at galactic
scales $K_0\sim \mathcal{O}(400-7000)$$(km/sec)^2$. This also fits
the galaxies presented in the previous section, which gave an
optimal $K_0$ of the order $\mathcal{O}(100)$$(km/sec)^2$. Thus,
the scale-dependent EoS DM seems to be described by the universal
law,
\begin{equation}\label{univrersallaw}
\boxed{\langle P(\chi)\rangle =\langle \rho(\chi)\rangle \langle
\sigma_{DM}^2\rangle \, ,}
\end{equation}
where $\chi$ is the characteristic scale under consideration. The
characteristic velocity dispersion is something that is
dynamically adjustable to the background considered, and the
entropy function $K(\chi)$ is related to it by,
\begin{equation}\label{universallaw1}
\boxed{\langle K(\chi)\rangle=\langle \sigma_{DM}^2\rangle}\, .
\end{equation}
At cosmological scales, for distances $\chi \geq 100$Mpc, the
background is homogeneous, and the coarse-grained function
$K(\chi)$ takes a universal value, while at galactic scales, the
background is inhomogeneous, dynamical and is affected by a
non-linear gravitational bound structure, the galaxy. In each
galaxy, the characteristic velocity dispersion, and thus the
entropy function $K(r)$ is affected locally and dynamically by the
gravitational potential, the halo formation history, the process
of violent relaxation, the continuous accretion and the baryonic
gravitational potential. This supports a radius dependence in the
velocity dispersion and thus in the entropy function $K(r)$, which
is absent at cosmological scales. At the outer halo, the function
$K(r)$ approaches a constant because the physics are simplified at
larger distances from the galactic core, while the inner halo is
more physically complicated for the reasons we quoted above. The
flatness of the galactic rotation curves require a constant $K(r)$
at large distances from the core of the galaxy, while a simple
constant $K(r)$ cannot capture correctly the behavior near the
bulge, thus a radius dependent profile is optimal for fitting the
galactic rotation curves, and thus even solving the cusp-core
problem for DM dominated galaxies, like for example dwarfs and low
luminosity or low surface brightness spirals.

Now, the value of $w_{DM_0}$ at cosmological scales is somewhat
important because we need to have an order of magnitude estimation
for this, for cosmological phenomenology reasons. The same applies
at galactic scales. At galactic scales, $\sigma_{DM}\sim
100-300$$Km/sec$ thus from relation
(\ref{characteristicvelocityandKr}), this yields $w_{DM}\sim
10^{-7}$, where $w_{DM}=\frac{\partial P}{\partial
\rho}=K(\chi)/c^2$. On cosmological scales, the coarse grained
velocity dispersion is by far smaller, since from relations
(\ref{characteristicvelocityandKr}) and (\ref{eoseparametercosmo})
and since at cosmological scales we have $\sigma_{DM}^2\sim
\mathcal{O}(1)$$(Km/sec)^2$, we get, $w_{DM_0}\sim 10^{-11}$ at
small redshifts. This is extremely small, and thus at cosmological
scales this sort of DM is indistinguishable from cold DM. In fact,
even at recombination redshifts, one has $w_{DM}\sim 10^{-5}$.
However, there are issues at larger redshifts that may affect the
clustering of the matter perturbations due to the non-trivial
cosmological sound speed and the effects on the Jeans length. More
importantly, there is no particular reason that the DM we are
discussing so far is the only component of DM that can contribute
to the cosmological abundance of DM cosmologically and thus to the
matter perturbations and the sound speed. These features are
somewhat important and we need to put them into proper context. We
thoroughly discuss these issues in the following section.

\section{Matter Perturbations, Clustering and the Cosmological Jeans Scale}

Assuming that the sole component of DM is the scale-dependent EoS
DM, the behavior of the sound speed in Eq.
(\ref{soundspeedcosmol1}) may have important implications for the
evolution of matter perturbations during the matter domination
epoch. The matter perturbations evolution equation is,
\begin{equation}\label{matterperturbationevo}
\ddot{\delta}_m+2H\dot{\delta}_m\left(\frac{k^2c_s^2}{a^2}-4\pi G
\rho_{DM} \right)\delta_m=0\, ,
\end{equation}
where $\delta_m=\frac{d\rho_{DM}}{\rho_DM}$. The corresponding
Jeans wavenumber $k_J$ is,
\begin{equation}\label{jeanswavenumber}
k_J\sim \frac{a\sqrt{4\pi G \rho_{DM}}}{c_s}\, .
\end{equation}
In the above context, from Eq. (\ref{soundspeedcosmol1}) we have
$c_s\sim a^{-1}$, thus $k_J\sim a^{1/2}$. Now the Jeans length of
this size and the corresponding value of the sound speed could
seriously affect the clustering and the evolution of the
perturbations during the matter and the radiation epochs.

However, there is a subtlety in our syllogism which we now briefly
discuss, and also an alternative viewpoint for DM and its
components as a whole. With regard to the latter, DM might be
composed by components which actually contribute differently at
galactic and cosmological scales. This is a new perspective we
would like to introduce briefly, after we explain the subtlety
with the sound speed and the Jeans scale.

For our derivation of the sound speed in Eq.
(\ref{soundspeedcosmol1}), we used the rest frame effective sound
speed definition, which is defined via the pressure and density
perturbations in the DM rest frame,
\begin{equation}\label{restframe}
c_{eff}^2=\left( \frac{\delta P_{DM}}{\delta
\rho_{DM}}\right)_{rest}\, ,
\end{equation}
which coincides with $c_s^2=\frac{\partial P}{\partial \rho}$ only
for a barotropic fluid $P(\rho)$. But the scale-dependent EoS DM
is not a barotropic fluid at cosmological scales. In fact, the EoS
for DM at cosmological scales, namely Eq.
(\ref{globalScaleDependentEoSaverage}), is an EoS of the form
$P=P(a,\rho)$ where $a$ is the scale factor. The standard and
conventional definition of the adiabatic sound speed in a
homogeneous background is,
\begin{equation}\label{adiabaticsoundspeed}
c_a^2=\frac{\dot{\bar{P}}_{DM}}{\dot{\bar{\rho}}_{DM}}\, ,
\end{equation}
and in our case we have $\bar{P}_{DM}=K(a)\bar{\rho}_{DM}$, with
$K(a)=K_0a^{-2}$ and thus $w_{DM}(a)=K_0a^{-2}$. Since DM is a
perfect fluid, the conservation of energy equation
(\ref{contequation}) implies that the adiabatic sound speed is,
\begin{equation}\label{adiabaticsoudnspeeddef}
c_a^2=w_{DM}(a)-\frac{\dot{w}_{DM}(a)}{3H(1+w_{DM}(a))}\, ,
\end{equation}
and since $w_{DM}\sim a^{-2}$ we get,
\begin{equation}\label{wdmfinaladiabaticsoundspeed}
\frac{c_a^2}{c^2}=w_{DM}(a)+\frac{2w_{DM}(a)}{3(1+w_{DM}(a))}\, .
\end{equation}
So the adiabatic sound speed is not identical to the barotropic
sound speed at rest. Now this sound speed can also deviate from
the standard sound speed of cold DM and this can strongly affect
the evolution of the matter perturbations and the Jeans scale
during matter domination era. The complete and actually formal
approach would then be to solve the linearized Einstein equations
together with the scale-dependent EoS DM using a Boltzmann solver
and compare the results with the cold DM standard perturbations.
One must use identical initial conditions and correctly chosen
background cosmological parameters. Then we can extract the DM
density contrast $\delta_{DM}(k,a)=\frac{\delta
\rho_{DM}(k,a)}{\bar{\rho}_{DM}(k,a)}$, and also the difference
between the cold DM and the scale-dependent EoS DM quantified in
the form of a transfer function
$T(k,a)=\frac{\delta_{DM}^{model}(k,a)}{\delta_{CDM}(k,a)}$ and
also study the evolution of Jeans scale.

There is however another subtlety, since $w_{DM}$ redshifts as
$w_{DM}\sim a^{-2}$ which can be fatal at large redshifts. This is
however an artifact of the scaling $K(\chi)\sigma_{DM}^2$, so
basically the EoS is scaling as the characteristic velocity
dispersion $w_{DM}\sim \sigma_{DM}^2$. For non-relativistic
particles $\sigma_{DM}^2\sim a^{-2}$, but as the redshift
increases, the law $\sigma_{DM}^2\sim a^{-2}$ ceases to hold true.
So basically $w_{DM}\sim a^{-2}$ holds true when the
scale-dependent EoS DM is non-relativistic. In general, for an
isotropic and collisionless DM, with distribution function $f(q)$,
with $q=ap$ being the co-moving conserved momentum, the energy
density and pressure are given by,
\begin{equation}\label{energy}
\rho(a)=\frac{g}{2\pi^2a^4}\int_{0}^{\infty}\mathrm{d}q\,q^2\sqrt{q^2+a^2m^2}f(q)\,
,
\end{equation}
and
\begin{equation}\label{pressurefq}
P(a)=\frac{g}{6\pi^2a^4}\int_{0}^{\infty}\mathrm{d}q\frac{q^4}{\sqrt{q^2+a^2m^2}}f(q)\,
,
\end{equation}
where $g$ denotes the internal degrees of freedom of the DM
particles. For $am\ll q$, one gets $P=\rho/3$ and $w_{DM}=1/3$
while for $am\gg q$ one gets $w_{DM}\sim a^{-2}$. Thus
interpolating the relation $w_{DE}\sim a^{-2}$ arbitrarily in the
past makes no sense, because it is not valid. This feature should
also be taken into account when comparing the perturbations of
ordinary cold DM and that of scale-dependent EoS DM.

But let us discuss another perspective regarding DM, the central
theme of which is in a nutshell that the DM which participates in
galactic dynamics might not be the same component of DM which
participates at the cosmological dynamics. Specifically, the total
DM might be comprised by the DM that participates at galactic
dynamics phenomena, and also there can be another component which
dominates the cosmological energy density, so we actually have,
\begin{equation}\label{totaldmcomponents}
\rho_{DM}^{tot}=\rho_{gal}+\rho_X\, ,
\end{equation}
where $\rho_{gal}$ is the galactic DM component and $\rho_X$ is
the the DM component which participates at galactic phenomena. The
galactic component, which is subdominant cosmologically, at
galactic scales is described by the virialized EoS
$P(r)=K(r)\rho(r)$ while at cosmological scales, in the
non-relativistic limit, the galactic DM EoS behaves as $P(a)=K_0
a^{-2}\rho(a)$, with the sound speed and the DM EoS parameter as
we discussed above. This DM component has specific clustering
behavior, which is not identical to the clustering properties of
the component $\rho_X$. The latter can originate from a DM sector
which has an alternative velocity dispersion, or even have
wave-like dynamics (Bose-Einstein condensates for example) or even
be sterile neutrinos, which do not cluster. These DM sector may
have an inherent suppression scale and such inherent mechanisms
may allow for a component which although contributes at
cosmological scales and dynamics, do not contribute to galactic
scale dynamics and thus this component is prevented from
clustering at sufficiently small scales.

This multi-component perspective of DM is rather interesting
because the dark component which does not cluster may have
interesting consequences. For example its abundance cosmologically
and its abundance inside a virialized DM halo are not identically
related. This component may contribute significantly
cosmologically, while it remains dynamically subdominant at
virialized halos, if for example its perturbations are suppressed
at galactic scales.

The two DM fluids may be perfect fluids, or even interacting
fluids, so their continuity equations behave in general as follows
in a FRW background,
\begin{align}\label{coupledfluidsDM}
& \dot{\rho}_{gal}+3H(\rho_{gal}+P_{gal})=-Q(H,\rho_{gal},\rho_X)\, ,\\
\notag &
\dot{\rho}_{X}+3H(\rho_{X}+P_{X})=Q(H,\rho_{gal},\rho_X)\, ,
\end{align}
so the scale-dependent EoS DM fluid might lose energy
cosmologically, to the other DM fluid. In both the non-interacting
and interacting cases of the DM fluids, the two components have
background evolutions and perturbations which are coupled,
therefore the whole discussion about the cosmological DM EoS
parameter and the sound speed may be embedded in an entirely new
context. The same behavior is expected in the case without
interaction between the two DM fluids, except their gravitational
interaction. The latter case is relatively more simple though.
Regarding the possible candidate theories for the dark sector
$\rho_X$, there are two main theories as we mentioned earlier,
massive neutrinos and ultra-light scalar fields of Bose-Einstein
condensate types. The neutrinos have thermal velocities which
generate free streaming and thus the growth of perturbations are
suppressed below a scale, but the allowed neutrino abundance is
small to be considered for a large amount of contribution in the
total DM fluid. The ultralight scalar perspective on the other is
more appealing because in such theories there are gradient wave
effects which can generate scale-dependent pressure responses.
Also the mass of the scalar can crucial affect the clustering
properties. These non-trivial pressure responses and the
appropriate mass of the scalar, can suppress clustering at small
scales, while at large scales the fluid behaves as pressure-less
matter.

Therefore, the cosmological DM abundance is not necessarily
related with a specific clustering behavior at all scales. This
simply means that the scale-dependent EoS DM could co-exist with
another DM component which is cosmologically dominant and the
clustering properties of which are suppressed at galactic scales.
This multi-component perspective, also changes the Jeans scale and
sound speed discussion, because cosmologically the dominant
component affects the Jeans scale and the sound speed and matter
density perturbations more efficiently. The same applies with the
CMB properties, such as the relevant baryon acoustic oscillations
and so on. Hence if the suppression scale of the cosmological
dominant component is large than the galactic scales of interest,
the contribution of this fluid on the formation of galactic halos
will be reduced, but the scale-dependent EoS DM may contribute
significantly at these scales. This two-component or
multi-component DM which distinguishes between DM species that
control the galactic dynamics and ones which control cosmological
scales is not new and it is interesting in its own right. And it
aligns with various observations which distinguish DM behaviors,
for example the Bullet cluster and the Abell 520 supercluster. In
the literature there are relevant works of two-component DM
fluids, which can interact, and in some cases these multi-fluids
interacting between them can explain galactic dynamics problems of
the $\Lambda$CDM, see
\cite{Todoroki:2017kge,Kim:2023onk,Duan:2026zqj}. In the same line
of research, one may add an interaction between one of the two DM
components and dark energy. Such models have recently been
introduced to address the latest DESI observations
\cite{DiValentino:2020naf,Dai:2020rfo,He:2020zns,Nakai:2020oit,Agrawal:2019dlm,Yang:2018euj,
Ye:2020btb,Vagnozzi:2021tjv,Desmond:2019ygn,Hogas:2023pjz,OColgain:2018czj,Vagnozzi:2019ezj,
Krishnan:2020obg,Colgain:2019joh,Vagnozzi:2021gjh,Lee:2022cyh,Krishnan:2021dyb,Ye:2021iwa,Ye:2022afu,Verde:2019ivm,Menci:2024rbq,Adil:2023ara,Reeves:2022aoi,Ferlito:2022mok,
Vagnozzi:2021quy,DiValentino:2020evt,DiValentino:2019ffd,DiValentino:2025sru,Valletta:2025bgu,Montani:2025nmz,Fazzari:2025mww,Montani:2025jkk,Schiavone:2024heb,Montani:2024pou,Montani:2024ntj,Escamilla:2024xmz,Montani:2023ywn,Montani:2023xpd}.

\section{Conclusions}

We studied the formal extrapolation of scale-dependent EoS DM from
galactic to cosmological scales. The galactic scale-dependent EoS
DM profile which can optimally fit the SPARC data was found from
previous works to be a nearly isothermal EoS DM of the form
$P(r)=K(r)\rho(r)$, where $K(r)$ is an entropy function with a
radial dependence and it also must be slowly varying.  At
cosmological scales, this galactic DM profile is generalized as
$\langle P(\chi)\rangle =\langle \rho(\chi)\rangle \langle
\sigma_{DM}^2\rangle$, where $\chi$ is the scale of the system
under study and $\sigma_{DM}$ is the characteristic velocity
dispersion of DM at the corresponding scale. Thus the entropy
function is generalized to be $K(\chi)=\langle
\sigma_{DM}^2\rangle$. We studied the behavior of this sort of DM
at cosmological scales, and for consistency it is found that this
matter can describe galactic dynamics solely, but it cannot be the
sole component of DM in the Universe, since another dominant
component must exist who can control the cosmological evolution.
We also introduced the two-component DM concept, where the
scale-dependent EoS DM can only control the galactic structures,
and the other component can only control the cosmological scales.
The two DM components must have different clustering properties
and we briefly discussed the physics of the two component DM
concept. It is interesting to study the two-component DM scenario
at a cosmological context, with a non-trivial interaction between
the two components, and even combine these two with the dark
energy fluid with or without interaction with one of the DM
fluids. This task is useful in view of the DESI data on dark
energy.


\begin{thebibliography}{99}



\bibitem{DiValentino:2020naf}
E.~Di Valentino, A.~Mukherjee and A.~A.~Sen,
Entropy \textbf{23} (2021) no.4, 404 doi:10.3390/e23040404
[arXiv:2005.12587 [astro-ph.CO]].

\bibitem{Dai:2020rfo}
W.~M.~Dai, Y.~Z.~Ma and H.~J.~He,
Phys. Rev. D \textbf{102} (2020), 121302
doi:10.1103/PhysRevD.102.121302 [arXiv:2003.03602 [astro-ph.CO]].

\bibitem{He:2020zns}
H.~J.~He, Y.~Z.~Ma and J.~Zheng,
JCAP \textbf{11} (2020), 003 doi:10.1088/1475-7516/2020/11/003
[arXiv:2003.12057 [hep-ph]].

\bibitem{Nakai:2020oit}
Y.~Nakai, M.~Suzuki, F.~Takahashi and M.~Yamada,
Phys. Lett. B \textbf{816} (2021), 136238,
[astro-ph.CO]].


\bibitem{Agrawal:2019dlm}
P.~Agrawal, G.~Obied and C.~Vafa,
Phys. Rev. D \textbf{103} (2021) no.4, 043523, 
[arXiv:1906.08261 [astro-ph.CO]].

\bibitem{Yang:2018euj}
W.~Yang, S.~Pan, E.~Di Valentino, R.~C.~Nunes, S.~Vagnozzi and
D.~F.~Mota,
JCAP \textbf{09} (2018), 019, 
[arXiv:1805.08252 [astro-ph.CO]].

\bibitem{Ye:2020btb}
G.~Ye and Y.~S.~Piao,
Phys. Rev. D \textbf{101} (2020) no.8, 083507,
 [arXiv:2001.02451 [astro-ph.CO]].

\bibitem{Vagnozzi:2021tjv}
S.~Vagnozzi, F.~Pacucci and A.~Loeb,
JHEAp \textbf{36} (2022), 27-35, 
[arXiv:2105.10421 astro-ph.CO]].

\bibitem{Desmond:2019ygn}
H.~Desmond, B.~Jain and J.~Sakstein,
Phys. Rev. D \textbf{100} (2019) no.4, 043537 [erratum: Phys. Rev.
D \textbf{101} (2020) no.6, 069904; erratum: Phys. Rev. D
\textbf{101} (2020) no.12, 129901] doi:10.1103/PhysRevD.100.043537
[arXiv:1907.03778 [astro-ph.CO]].

\bibitem{Hogas:2023pjz}
M.~H\"og\r{a}s and E.~M\"ortsell,
Phys. Rev. D \textbf{108} (2023) no.12, 124050
doi:10.1103/PhysRevD.108.124050 [arXiv:2309.01744 [astro-ph.CO]].

\bibitem{OColgain:2018czj}
E.~\'O Colg\'ain, M.~H.~P.~M.~van Putten and H.~Yavartanoo,
Phys. Lett. B \textbf{793} (2019), 126-129, 
[arXiv:1807.07451 [hep-th]].

\bibitem{Vagnozzi:2019ezj}
S.~Vagnozzi,
Phys. Rev. D \textbf{102} (2020) no.2, 023518, 
[arXiv:1907.07569 [astro-ph.CO]].

\bibitem{Krishnan:2020obg}
C.~Krishnan, E.~\'O.~Colg\'ain, Ruchika, A.~A.~Sen,
M.~M.~Sheikh-Jabbari and T.~Yang,
Phys. Rev. D \textbf{102} (2020) no.10, 103525
[arXiv:2002.06044 [astro-ph.CO]].

\bibitem{Colgain:2019joh}
E.~\'O.~Colg\'ain and H.~Yavartanoo,
Phys. Lett. B \textbf{797} (2019), 134907
doi:10.1016/j.physletb.2019.134907 [arXiv:1905.02555
[astro-ph.CO]].

\bibitem{Vagnozzi:2021gjh}
S.~Vagnozzi,
Phys. Rev. D \textbf{104} (2021) no.6, 063524
doi:10.1103/PhysRevD.104.063524 [arXiv:2105.10425 [astro-ph.CO]].

\bibitem{Lee:2022cyh}
B.~H.~Lee, W.~Lee, E.~\'O.~Colg\'ain, M.~M.~Sheikh-Jabbari and
S.~Thakur,
JCAP \textbf{04} (2022) no.04, 004, 
 [arXiv:2202.03906 [astro-ph.CO]].


\bibitem{Krishnan:2021dyb}
C.~Krishnan, R.~Mohayaee, E.~\'O.~Colg\'ain, M.~M.~Sheikh-Jabbari
and L.~Yin,
Class. Quant. Grav. \textbf{38} (2021) no.18, 184001
doi:10.1088/1361-6382/ac1a81 [arXiv:2105.09790 [astro-ph.CO]].


\bibitem{Ye:2021iwa}
G.~Ye, J.~Zhang and Y.~S.~Piao,
[arXiv:2107.13391 [astro-ph.CO]].

\bibitem{Ye:2022afu}
G.~Ye and Y.~S.~Piao,
Phys. Rev. D \textbf{106} (2022) no.4, 043536, 
[arXiv:2202.10055 [astro-ph.CO]].


\bibitem{Verde:2019ivm}
L.~Verde, T.~Treu and A.~G.~Riess,
Nature Astron. \textbf{3}, 891 doi:10.1038/s41550-019-0902-0
[arXiv:1907.10625 [astro-ph.CO]].


\bibitem{Menci:2024rbq}
N.~Menci, S.~A.~Adil, U.~Mukhopadhyay, A.~A.~Sen and S.~Vagnozzi,
JCAP \textbf{07} (2024), 072, 
[arXiv:2401.12659 [astro-ph.CO]].

\bibitem{Adil:2023ara}
S.~A.~Adil, U.~Mukhopadhyay, A.~A.~Sen and S.~Vagnozzi,
JCAP \textbf{10} (2023), 072, 
 [arXiv:2307.12763 [astro-ph.CO]].

\bibitem{Reeves:2022aoi}
A.~Reeves, L.~Herold, S.~Vagnozzi, B.~D.~Sherwin and
E.~G.~M.~Ferreira,
Mon. Not. Roy. Astron. Soc. \textbf{520} (2023) no.3, 3688-3695, 
[arXiv:2207.01501 [astro-ph.CO]].

\bibitem{Ferlito:2022mok}
F.~Ferlito, S.~Vagnozzi, D.~F.~Mota and M.~Baldi,
Mon. Not. Roy. Astron. Soc. \textbf{512} (2022) no.2, 1885-1905, 
[arXiv:2201.04528 [astro-ph.CO]].

\bibitem{Vagnozzi:2021quy}
S.~Vagnozzi, L.~Visinelli, P.~Brax, A.~C.~Davis and J.~Sakstein,
Phys. Rev. D \textbf{104} (2021) no.6, 063023, 
[arXiv:2103.15834 [hep-ph]].

\bibitem{DiValentino:2020evt}
E.~Di Valentino, S.~Gariazzo, O.~Mena and S.~Vagnozzi,
JCAP \textbf{07} (2020) no.07, 045. 
[arXiv:2005.02062[astro-ph.CO]].


\bibitem{DiValentino:2019ffd}
E.~Di Valentino, A.~Melchiorri, O.~Mena and S.~Vagnozzi,
Phys. Dark Univ. \textbf{30} (2020), 100666, 
[arXiv:1908.04281 [astro-ph.CO]].

\bibitem{DiValentino:2025sru}
E.~Di Valentino, J.~Levi Said, A.~Riess, A.~Pollo, V.~Poulin,
A.~G\'omez-Valent, A.~Weltman, A.~Palmese, C.~D.~Huang and C.~van
de Bruck, \textit{et al.}
[arXiv:2504.01669 [astro-ph.CO]].


\bibitem{Valletta:2025bgu}
A.~Valletta, G.~Montani, M.~G.~Dainotti and E.~Fazzari,
[arXiv:2512.19568 [gr-qc]].


\bibitem{Montani:2025nmz}
G.~Montani, L.~A.~Escamilla, N.~Carlevaro and E.~Di Valentino,
Phys. Rev. D \textbf{113} (2026) no.2, 023507
doi:10.1103/mn69-1dn6 [arXiv:2512.20193 [astro-ph.CO]].




\bibitem{Fazzari:2025mww}
E.~Fazzari, M.~G.~Dainotti, G.~Montani and A.~Melchiorri,
JHEAp \textbf{49} (2026), 100459 doi:10.1016/j.jheap.2025.100459
[arXiv:2506.04162 [astro-ph.CO]].



\bibitem{Montani:2025jkk}
G.~Montani, M.~De Angelis and M.~G.~Dainotti,
Phys. Dark Univ. \textbf{49} (2025), 101969
doi:10.1016/j.dark.2025.101969 [arXiv:2506.13288 [astro-ph.CO]].



\bibitem{Schiavone:2024heb}
T.~Schiavone and G.~Montani,
Nuovo Cim. C \textbf{48} (2025) no.3, 105
doi:10.1393/ncc/i2025-25105-3 [arXiv:2408.01410 [gr-qc]].


\bibitem{Montani:2024pou}
G.~Montani, N.~Carlevaro, L.~A.~Escamilla and E.~Di Valentino,
Phys. Dark Univ. \textbf{48} (2025), 101848
doi:10.1016/j.dark.2025.101848 [arXiv:2404.15977 [gr-qc]].



\bibitem{Montani:2024ntj}
G.~Montani, N.~Carlevaro and M.~G.~Dainotti,
Phys. Dark Univ. \textbf{48} (2025), 101847
doi:10.1016/j.dark.2025.101847 [arXiv:2411.07060 [gr-qc]].


\bibitem{Escamilla:2024xmz}
L.~A.~Escamilla, D.~Fiorucci, G.~Montani and E.~Di Valentino,
Phys. Dark Univ. \textbf{46} (2024), 101652
doi:10.1016/j.dark.2024.101652 [arXiv:2408.04354 [astro-ph.CO]].



\bibitem{Montani:2023ywn}
G.~Montani, N.~Carlevaro and M.~G.~Dainotti,
Phys. Dark Univ. \textbf{44} (2024), 101486
doi:10.1016/j.dark.2024.101486 [arXiv:2311.04822 [gr-qc]].



\bibitem{Montani:2023xpd}
G.~Montani, M.~De Angelis, F.~Bombacigno and N.~Carlevaro,
Mon. Not. Roy. Astron. Soc. \textbf{527} (2023) no.1, L156-L161
doi:10.1093/mnrasl/slad159 [arXiv:2306.11101 [gr-qc]].






\bibitem{Tulin:2017ara}
S.~Tulin and H.~B.~Yu,
Phys. Rept. \textbf{730} (2018), 1-57
doi:10.1016/j.physrep.2017.11.004 [arXiv:1705.02358 [hep-ph]].



\bibitem{Oikonomou:2026zhu}
V.~K.~Oikonomou,
Phys. Dark Univ. \textbf{54} (2026), 102449
doi:10.1016/j.dark.2026.102449


\bibitem{Oikonomou:2026seq}
V.~K.~Oikonomou and E.~I.~Manouri,
Phys. Dark Univ. \textbf{54} (2026), 102439
doi:10.1016/j.dark.2026.102439 [arXiv:2609.29495 [gr-qc]].


\bibitem{Oikonomou:2025bsi}
V.~K.~Oikonomou,
Phys. Dark Univ. \textbf{52} (2026), 102346
doi:10.1016/j.dark.2026.102346 [arXiv:2511.13929 [astro-ph.CO]].



\bibitem{Oikonomou:2026vkp}
V.~K.~Oikonomou,
[arXiv:2601.20070 [gr-qc]].





\bibitem{Todoroki:2017kge}
K.~Todoroki and M.~V.~Medvedev,
Mon. Not. Roy. Astron. Soc. \textbf{483} (2019) no.3, 4004-4019
doi:10.1093/mnras/sty3353 [arXiv:1711.11085 [astro-ph.CO]].


\bibitem{Kim:2023onk}
J.~H.~Kim, K.~Kong, S.~H.~Lim and J.~C.~Park,
PTEP \textbf{2024} (2024) no.12, 123B03 doi:10.1093/ptep/ptae169
[arXiv:2312.07660 [hep-ph]].


\bibitem{Duan:2026zqj}
X.~C.~Duan, Y.~L.~S.~Tsai and Z.~Wang,
JCAP \textbf{11} (2026), 028 doi:10.1088/1475-7516/2026/11/028
[arXiv:2604.15006 [astro-ph.CO]].
\bibitem{Slepian:2011ev}
Z.~Slepian and J.~Goodman,
Mon. Not. Roy. Astron. Soc. \textbf{427} (2012), 839
doi:10.1111/j.1365-2966.2012.21901.x [arXiv:1109.3844
[astro-ph.CO]].






\bibitem{Saxton:2014swa}
C.~J.~Saxton, R.~Soria and K.~Wu,
Mon. Not. Roy. Astron. Soc. \textbf{445} (2014) no.4, 3415-3434
doi:10.1093/mnras/stu1984 [arXiv:1409.6725 [astro-ph.GA]].


\bibitem{Alonso-Alvarez:2024gdz}
G.~Alonso-{\'A}lvarez, J.~M.~Cline and C.~Dewar,
Phys. Rev. Lett. \textbf{133} (2024) no.2, 021401
doi:10.1103/PhysRevLett.133.021401 [arXiv:2401.14450
[astro-ph.CO]].



\bibitem{Kaplinghat:2015aga}
M.~Kaplinghat, S.~Tulin and H.~B.~Yu,
Phys. Rev. Lett. \textbf{116} (2016) no.4, 041302
doi:10.1103/PhysRevLett.116.041302 [arXiv:1508.03339
[astro-ph.CO]].


\bibitem{Ahn:2004xt}
K.~J.~Ahn and P.~R.~Shapiro,
Mon. Not. Roy. Astron. Soc. \textbf{363} (2005), 1092-1124
doi:10.1111/j.1365-2966.2005.09492.x [arXiv:astro-ph/0412169
[astro-ph]].



\bibitem{Saxton:2012ja}
C.~J.~Saxton,
Mon. Not. Roy. Astron. Soc. \textbf{430} (2013), 1578
doi:10.1093/mnras/sts689 [arXiv:1212.5294 [astro-ph.CO]].


\bibitem{Saxton:2010jk}
C.~J.~Saxton and I.~Ferreras,
Mon. Not. Roy. Astron. Soc. \textbf{405} (2010), 77
doi:10.1111/j.1365-2966.2010.16448.x [arXiv:1002.0845
[astro-ph.CO]].


\bibitem{Saxton:2016ozz}
C.~J.~Saxton, Z.~Younsi and K.~Wu,
Mon. Not. Roy. Astron. Soc. \textbf{461} (2016) no.4, 4295-4316
doi:10.1093/mnras/stw1626 [arXiv:1606.07066 [astro-ph.GA]].


\bibitem{Arbey:2003sj}
A.~Arbey, J.~Lesgourgues and P.~Salati,
Phys. Rev. D \textbf{68} (2003), 023511
doi:10.1103/PhysRevD.68.023511 [arXiv:astro-ph/0301533
[astro-ph]].


\bibitem{Heikinheimo:2015kra}
M.~Heikinheimo, M.~Raidal, C.~Spethmann and H.~Veerm{\"a}e,
Phys. Lett. B \textbf{749} (2015), 236-241
doi:10.1016/j.physletb.2015.08.012 [arXiv:1504.04371 [hep-ph]].









\bibitem{Wandelt:2000ad}
B.~D.~Wandelt, R.~Dave, G.~R.~Farrar, P.~C.~McGuire, D.~N.~Spergel
and P.~J.~Steinhardt,
[arXiv:astro-ph/0006344 [astro-ph]].






\bibitem{Spergel:1999mh}
D.~N.~Spergel and P.~J.~Steinhardt,
Phys. Rev. Lett. \textbf{84} (2000), 3760-3763
doi:10.1103/PhysRevLett.84.3760 [arXiv:astro-ph/9909386
[astro-ph]].





\bibitem{Loeb:2010gj}
A.~Loeb and N.~Weiner,
Phys. Rev. Lett. \textbf{106} (2011), 171302
doi:10.1103/PhysRevLett.106.171302 [arXiv:1011.6374
[astro-ph.CO]].


\bibitem{Ackerman:2008kmp}
L.~Ackerman, M.~R.~Buckley, S.~M.~Carroll and M.~Kamionkowski,
Phys. Rev. D \textbf{79} (2009), 023519
doi:10.1103/PhysRevD.79.023519 [arXiv:0810.5126 [hep-ph]].



\bibitem{Goodman:2000tg}
J.~Goodman,
New Astron. \textbf{5} (2000), 103
doi:10.1016/S1384-1076(00)00015-4 [arXiv:astro-ph/0003018
[astro-ph]].



\bibitem{Arbey:2006it}
A.~Arbey,
Phys. Rev. D \textbf{74} (2006), 043516
doi:10.1103/PhysRevD.74.043516 [arXiv:astro-ph/0601274
[astro-ph]].



\bibitem{Yang:2025ume}
W.~Yang, S.~Pan, E.~Di Valentino, O.~Mena, D.~F.~Mota and
S.~Chakraborty,
Phys. Rev. D \textbf{111} (2025) no.10, 103509
doi:10.1103/PhysRevD.111.103509 [arXiv:2504.11973 [astro-ph.CO]].


\bibitem{Abedin:2025dis}
M.~Abedin, L.~A.~Escamilla, S.~Pan, E.~Di Valentino and W.~Yang,
[arXiv:2505.09470 [astro-ph.CO]].



\bibitem{Nunes:2026wfg}
R.~C.~Nunes,
[arXiv:2611.08636 [astro-ph.CO]].



\bibitem{refugca320}

Qinwei Yuan, Yong Shi, Xin Li, Zhiyu Zhang, Zhiyuan Zheng,
Qiusheng Gu, Xiaoling Yu, Fuyan Bian, arXiv:2611.15528




\end{thebibliography}
\end{document}